\documentstyle[12pt]{article}
\makeatletter

\@addtoreset{equation}{section}
\def\theequation{\arabic{section}.\arabic{equation}}

\def\section{\@startsection{section}{1}{\z@}{3.5ex plus 1ex minus
   .2ex}{2.3ex plus .2ex}{\large\bf}}
   \def\thesection{\arabic{section}}

\def\appendix{\setcounter{section}{0}
        \def\thesection{Appendix\ \Alph{section}}
        \def\theequation{\Alph{section}.\arabic{equation}}}

\newcommand{\beq}{\begin{eqnarray}}
\newcommand{\eeq}{\end{eqnarray}}
\newcommand{\eq}{eqnarray}
\newcommand{\bb}{\bibitem}
\newcommand{\al}{{\alpha}}
\newcommand{\be}{{\beta}}
\newcommand{\ci}{\cite}

\newcommand{\de}{{\delta}}
\newcommand{\De}{\Delta}

\newcommand{\no}{{\nonumber}}
\newcommand{\f}{\frac}
\newcommand{\eff}{\hbox{\scriptsize eff}}
\newcommand{\AdS}{\hbox{\scriptsize AdS}}
\newcommand{\BTZ}{\hbox{\scriptsize BTZ}}
\newcommand{\HMTZ}{\hbox{\scriptsize HMTZ}}
\newcommand{\BH}{\hbox{\scriptsize BH}}
\newcommand{\Euc}{Euclidean }

\begin{document}

\topmargin 0pt

\oddsidemargin -3.5mm

\headheight 0pt

\topskip 0mm
\addtolength{\baselineskip}{0.20\baselineskip}
\begin{flushright}
\end{flushright}
\begin{flushright}
hep-th/0403089
\end{flushright}
\vspace{0.5cm}
\begin{center}
    {\large \bf  Fate of Three-Dimensional Black Holes Coupled to a
    Scalar Field and the Bekenstein-Hawking Entropy}\end{center}
\vspace{0.5cm}
\begin{center}
Mu-In Park\footnote{~Electronic address: muinpark@yahoo.com} \\
{ Department of Physics, POSTECH, Pohang 790-784, Korea}\\
\end{center}
\vspace{0.5cm}
\begin{center}
    {\bf ABSTRACT}
\end{center}
Three-dimensional black holes coupled to a self-interacting scalar
field
is considered. It is known that its statistical entropy $a'~la$
Strominger does $not$ agree with the Bekenstein-Hawking (BH)
entropy. However I show that, by a careful treatment of the vacuum
state in the {\it canonical} ensemble with a fixed temperature,
which is the same as that of the BTZ black hole without the scalar
field, the BH entropy may be exactly produced by the Cardy's
formula. I discuss its several implications, including the fate of
black holes, no-scalar-hair theorems, stability, mirror black
holes, and higher-order corrections to the entropy.

 \vspace{3cm}
\begin{flushleft}

13 July 2004 \\
\end{flushleft}

\newpage

\section{Introduction}
Three-dimensional gravity with a negative cosmological constant
$\Lambda=-1/l^2$ has been considered as the most concrete example
of the AdS/CFT correspondence, which states, roughly speaking,
\begin{\eq}
\label{AdS/CFT}
 Z_{\AdS}\sim Z_{\partial(\AdS)}
\end{\eq}
between the partition functions $Z_{\AdS}$ on the bulk AdS space
and $Z_{\partial(\AdS)}$ on its boundary $\partial(AdS)$
\ci{Ahar:99}. Especially for the BTZ black hole \ci{Bana:92},
Strominger has shown that the correspondence (\ref{AdS/CFT}) is
``precise'', from the knowledge of the asymptotic isometry group
$SO(2,2)$, which generates the two copies of the Virasoro algebra
with the central charges \ci{Brow:86,Oh:98,
Bana:99}
\begin{\eq}
\label{c} {\bar c}=c=\f{3 l}{2 G}~,
\end{\eq}
and the Cardy's formula \ci{Card:86} for the density of states
\begin{\eq}
\label{S-BTZ}
 \mbox{ln} \rho \sim 2 \pi \sqrt{\f{c
(\Delta-c/24)}{6}}+2 \pi \sqrt{\f{{\bar c} ({\bar
\Delta}-\bar{c}/24) }{6}}~
\end{\eq}
when the AdS$_3$ vacuum, which corresponds to the mass eigenvalue
of $M=-1/8G$, is chosen \ci{Stro:98}: $Z_{\BTZ}=Z_{\partial
({\BTZ}) } \sim \mbox{exp} \{2 \pi r_+/4 G \}$. ($r_+$ is the
radius of the outer event horizon.) From the frequent appearance
of the BTZ black hole geometry in the many higher dimensional
black holes in string theory,
this result has been far reaching consequence in the higher
dimensional AdS/CFT also \ci{Hyun:97}.

Recently, the generalization of the Strominger's result to the
black holes coupled to a self-interacting scalar field have been
studied \ci{Nats:99,Henn:02,Gege:03}. There it is found that the
asymptotic symmetry group and the algebra of the canonical
generators remain the same as in pure gravity even though the
fall-off of the fields at spatial infinity is {\it slower} than
that of pure gravity such as the symmetry generators acquire
non-trivial pieces from the scalar field. However, strangely, it
is known that the statistical entropy from the Cardy's formula
does not agree with the Bekenstein-Hawking (BH) entropy
\begin{\eq}
S_{\BH}=\f{2 \pi r_+}{4 G},
\end{\eq}
such as the usual AdS/CFT correspondence (\ref{AdS/CFT}) is $not$
favored; this has been first observed by Natsuume et al
\ci{Nats:99} and later by Henneaux et al \ci{Henn:02} in a more
general context; but no proper explanation or resolution has been
provided so far.

In this paper, I address this issue and I show that by a careful
treatment of the vacuum state in the {\it canonical} ensemble with
a fixed temperature, which is the same as that of the BTZ black
hole without the scalar field, the BH entropy is exactly produced
by the Cardy's formula.

\section{Asymptotic Symmetry of Three-Dimensional Gravity Coupled
to a Scalar Field}

In this section I review the result of Ref.\ci{Henn:02} on the
asymptotic symmetry of three-dimensional gravity minimally coupled
to a scalar field. The associated action is given by
\begin{\eq}
I=\f{1}{ \pi G} \int d^3 x \sqrt{-g} \left[\f{R}{16}-\f{1}{2}
(\nabla \phi)^2 -V(\phi) \right].
\end{\eq}
With the asymptotic conditions of the pure gravity \ci{Brow:86},
the scalar field does not contribute to the asymptotic symmetry
generator if it decays rapidly enough, i.e., $\phi \sim
r^{-(1+\varepsilon )}$ \ci{Brow:86,Stro:98,Nats:99}. There exists,
however, a more general form of the asymptotic conditions with a
$slower$ fall-off \ci{Henn:02}
\begin{equation}
\phi =\frac{\chi }{r^{1/2}}-\alpha \frac{\chi ^{3}}{r^{3/2}}
+O(r^{-5/2}), \label{phi}
\end{equation}
\begin{equation}
\begin{array}{lll}
g_{rr}= \displaystyle \frac{l^{2}}{r^{2}}-\frac{4l^{2}\chi
^{2}}{r^{3}}+O(r^{-4}), & & \displaystyle g_{tt} =
-\frac{r^{2}}{l^{2}}+O(1),
\\[2mm] g_{tr} = O(r^{-2}), &  & \displaystyle g_{\varphi
\varphi } =  r^{2}+O(1), \\[1mm] g_{\varphi r} = O(r^{-2}), &  & \
\displaystyle g_{t\varphi } =O(1),
\end{array}
\label{fall-off}
\end{equation}
where $\chi =\chi (t,\varphi )$, and $\alpha $ is an arbitrary
constant. The asymptotic isometry, which is generated by the
following asymptotic Killing vectors
\begin{eqnarray}
\xi ^{t} &=&l\left[ T^{+}+T^{-}+\frac{l^{2}}{2r^{2}}(\partial
_{+}^{2}T^{+}+\partial _{-}^{2}T^{-})\right] +O(r^{-4})\;,  \nonumber \\
\xi ^{r} &=&-r(\partial _{+}T^{+}+\partial _{-}T^{-})+O(r^{-1})\;,
\label{Killing} \\
\xi ^{\varphi } &=&T^{+}-T^{-}-\frac{l^{2}}{2r^{2}}(\partial
_{+}^{2}T^{+}-\partial _{-}^{2}T^{-})+O(r^{-4})  \nonumber
\end{eqnarray}
is the same as in pure gravity \ci{Brow:86} [$2\partial_\pm =l
\f{\partial}{\partial t} \pm \f{\partial}{\partial \phi}$] and
$T^\pm (\f{t}{l} \pm \phi)$ generate two independent copies of the
Virasoro algebra at the {\it classical} level
\begin{\eq}
\label{Virasoro}
&& \{L_m , L_n\}=i(m-n)L_{m+n} +\f{ic}{12} m (m^2-1) \de_{m+n,0}, \no \\
&&\{\bar{L}_m, \bar{L}_n\}=i(m-n) \bar{L}_{m+n} +\f{ic}{12} m (m^2-1) \de_{m+n,0}, \no \\
&&\{L_m, {\bar L}_n \}=0
\end{\eq}
with the central charge (\ref{c}) as in the pure gravity; this is
a very non-trivial result since there are non-trivial
contributions in the Virasoro generators $L_m, \bar{L}_m$ from the
scalar field.

\section{Black Holes with a Scalar Field}

In this section, I consider the exact three-dimensional black hole
solutions with a $regular$ scalar hair of
Henneaux-Mart\'{i}nez-Troncosoo-Zanelli (HMTZ) \ci{Henn:02} and
recall a few of its salient features for our considerations,
especially on the decay of the black hole into the BTZ black hole,
which has been considered recently by Gegenberg et al
\ci{Gege:03}. Further details can be founded in
\ci{Henn:02,Gege:03}.

Exact solutions for which the metric and the scalar field satisfy
asymptotic conditions (\ref{phi}) and (\ref{fall-off}) can be
obtained for a particular one-parameter family of potentials of
the form
\begin{equation}
V_{\nu }(\phi )=-\frac{1}{8l^{2}}\left( \cosh ^{6}\phi +\nu \sinh
^{6}\phi \right) \;.
\label{V}
\end{equation}
For $\nu > -1$, there is a solution that describes a static and
circularly symmetric black hole, with a scalar field which is
regular everywhere, given by \footnote{For $\nu=0$ this solution
reduces to the one found in the conformal frame $( \hat{g}_{\mu
\nu }=(1-\hat{\phi}^{2})^{-2}g_{\mu \nu },~\hat{\phi}=\tanh \phi)$
\ci{Mart:96}.}

\begin{equation}
\phi ={\rm arctanh}\sqrt{\frac{B}{H(r)+B}}\;,  \label{Scalar}
\end{equation}
where $B$ is a non-negative integration constant
\begin{\eq}
B=\sqrt{\f{8 G l^2 M}{3(1 +\nu)}},
\end{\eq}
for the black hole mass $M$, and
\begin{\eq}
H(r)=\frac{1}{2}\left( r+\sqrt{r^{2}+4Br}\right) \;.
\end{\eq}
The metric reads
\begin{equation}
ds^{2}=-\left( \frac{H}{H+B}\right) ^{2}F(r)dt^{2}+\left(
\frac{H+B}{H+2B} \right) ^{2}\frac{dr^{2}}{F(r)}+r^{2}d\varphi
^{2}\; \label{BTZ-scalar}
\end{equation}
with
\begin{\eq}
F=\frac{H^{2}}{l^{2}}-(1+\nu )\left(
\frac{3B^{2}}{l^{2}}+\frac{2B^{3}}{ l^{2}H}\right) \;.
\end{\eq}
This HMTZ's solution satisfy the asymptotic conditions (\ref{phi})
and (\ref{fall-off}) with $\alpha=2/3$. And the event horizon is
located at
\begin{\eq}
\label{horizon}
 r_{+}=l \Theta _{\nu }~ \sqrt{\f{8 G M}{3(1 +\nu)}}~,
\end{\eq}
where the constant $\Theta _{\nu }${\bf \ }is expressed, in terms
of $z=1+i\sqrt{\nu }$, as \ci{Schu:77}
\begin{\eq}
\Theta_{\nu}=2(z\bar{z})^{2/3}\frac{z^{2/3}-\bar{z}^{2/3}}{z-\bar{z}}~.
\end{\eq}

The BH entropy and the Hawking temperature are given by,
respectively,
\begin{\eq}
S_{\BH}&=&\frac{2 \pi r_{+}}{4 G}=  2 \pi l \sqrt{\f{M}{2 G}}~
\f{\Theta_{\nu}}{\sqrt{3 (1+\nu)}}~,
\label{SBH} \\
T&=&\f{\kappa}{2 \pi} =\frac{3(1+\nu )}{2 \pi l^2 \Theta _{\nu }^2
}~ r_+ \;. \label{T}
\end{\eq}
Note that the specific heat $C=\partial M/\partial T=\pi r_+/2G$,
which is exactly the same as that of the BTZ black hole with a
vanishing scalar field, is positive such as the canonical ensemble
with an equilibrium temperature with a heat bath exists. However,
for a fixed temperature, apart from the above HMTZ black hole
solution, the BTZ black hole can also be at equilibrium with the
heat bath. This gives, from $T^{\HMTZ}=T^{\BTZ}$ for the Hawking
temperature of the BTZ black hole $T^{\BTZ}=r_+^{\BTZ}/2 \pi l^2$
\ci{Bana:92}, the following relationship between the horizon radii
\begin{\eq}
\label{rr}
 r_{+}^{\HMTZ}=\frac{\Theta _{\nu }^{2}}{3(1+\nu
)}~r_{+}^{\BTZ}\;,
\end{\eq}
where $r_+^{\BTZ}$ is the horizon radius of the BTZ black hole
\ci{Gege:03}. In terms of the mass $M^{\HMTZ}$ of the HMTZ black
hole,
this becomes
\begin{\eq}
\label{MM}
\frac{M^{\HMTZ}}{M^{\BTZ}}=\frac{\Theta _{\nu }^{2}}{%
3(1+\nu )}<1\;,
\end{\eq}
where $M^{\BTZ}=r_+^2/(8 G l^2)$ is the mass of the BTZ black hole
and the inequality comes from $\Theta _{\nu }^2-{3(1+\nu )}<0$ for
$\nu >-1$. This implies that there is a non-vanishing probability
for the decay of the HMTZ black hole
into the BTZ black hole, induced by the thermal fluctuations: The
tunnelling amplitude is given by $\Gamma \approx A e^{- \Delta
I_{E}}$ \ci{Cole:77} with some determinant $A$ and the difference
between the \Euc actions of the higher and lower mass solutions at
the {\it same temperature} $\Delta I_{E}=I_{E}
[\bar{g}^{\BTZ}]-I[\bar{g}^{\HMTZ}]=\pi l \left[1 -\f{3(1+\nu)}{
\Theta^2_{\nu}} \right] \sqrt{\f{M^{\HMTZ} \Theta_{\nu}^2}{6 G (1
+\nu)}} ~>~0$, where $\bar{g}$ is the classical solution of the
Einstein equation.

\section{Statistical Entropy}

Once one has the Virasoro algebra of (\ref{Virasoro}), one can
compute its statistical entropy $S=\rm{ln} \rho$ for the density
of states $\rho$ as follows \ci{Card:86,Carl:99,Park:02}
\begin{\eq}
S= 2 \pi \sqrt{\f{c_{\eff} \Delta_{\eff}}{6}}+2 \pi \sqrt{\f{{\bar
c}_{\eff} {\bar \Delta}_{\eff} }{6}},
\end{\eq}
where
\begin{\eq}
&&c_{\eff} =c-24 \Delta_{\hbox{\scriptsize min}},~~ \Delta_{\eff}=\Delta-c/24, \no \\
&&\bar{c}_{\eff} =\bar{c}-24 \bar{\Delta}_{\hbox{\scriptsize
min}},~~ \bar{\Delta}_{\eff}=\bar{\Delta}-\bar{c}/24~,
\end{\eq}
and $\De_{\hbox{\scriptsize min}} (\bar{\De}_{\hbox{\scriptsize
min}})$ is the minimum of $L_0 (\bar{L_0})$ eigenvalue $\De
(\bar{\De})$; for $\De_{\hbox{\scriptsize min}}
=\bar{\De}_{\hbox{\scriptsize min}}=0$ this reduces to the formula
(\ref{S-BTZ}), which is the case for the BTZ black hole. Then, by
the usual relations \ci{Bana:99,Park:98}
\begin{\eq}
\Delta =\bar{\De} =\f{lM}{2} +\f{l}{16 G}
\end{\eq}
and the central charge that exactly agrees with (\ref{c}) of the
BTZ black hole, one finds that
[$\De_{\eff}=\bar{\De}_{\eff}=lM/2$]
\begin{\eq}
\label{S-M}
 S=4 \pi l \sqrt{{- M_{\hbox{\scriptsize min}}~ M}}~,
\end{\eq}
where $M_{\hbox{\scriptsize min}}$ is the minimum eigenvalue of
the HMTZ black hole mass $M$.

On the other hand, from the fact that, for a given Hawking
temperature, the BTZ black hole is more stable thermodynamically
than the HMTZ black hole,
the true vacuum would be determined by that of the BTZ black hole;
since the vacuum of the BTZ black hole is \footnote{ There are
some debates on this though. See for example
\ci{Carl:98,Bana:99b,Cho:99,Davi:00}. }
$M^{\BTZ}_{\hbox{\scriptsize min}}=-1/8 G$, let the
``corresponding'' minimum mass eigenvalue
$M^{\HMTZ}_{\hbox{\scriptsize min}}$ for the HMTZ black hole
is, by ``assuming'' the relation (\ref{MM}), given by
\begin{\eq}
\label{Mmin}
 M^{\HMTZ}_{\hbox{\scriptsize min}}=-\f{1}{8 G} \f{\Theta_{\nu}^2}{3 (1+\nu)}~,
\end{\eq}
such as
\begin{\eq}
\label{ceff}
 c_{\eff}=\left(\f{\Theta_{\nu}^2}{3 (1+\nu)} \right)
\f{3l}{2 G}~.
\end{\eq}
Then, by plugging (\ref{Mmin}) into (\ref{S-M}) one finally obtain
the BH entropy (\ref{SBH}) ``exactly''; in the previous
computations \ci{Nats:99,Henn:02} a naive vacuum of
$M^{\HMTZ}_{\hbox{\scriptsize min}}=-1/8G$, such as
$\Delta_{\hbox{\scriptsize min}}=\bar{\Delta}_{\hbox{\scriptsize
min}}=0$, was considered and as a result, a wrong entropy $S=2 \pi
l \sqrt{M/2G}$ was obtained; the true ground state of the HMTZ
black hole
is {\it lifted up} as $\Delta_{\hbox{\scriptsize
min}}=\bar{\Delta}_{\hbox{\scriptsize min}}=({l}/{16
G})(1-{\Theta_{\nu}^2}/{3 (1+\nu)} )(>0) $ by absorbing the scalar
field.

\section{Discussions}

1. For $0 \leq \nu \leq 1$ some abnormal phenomena happen: Another
horizon $r_{++}$ which looks like a cosmological horizon appears
outside the horizon $r_{+}$;
It is created at infinity for $\nu=0$; as $\nu$ increases from
$0$, it approaches to $r_{+}$ until meets $r_{+}$ for $\nu=1$;
$r_{++}$ is located inside of $r_{+}$ for $\nu > 1$ such as it
behaves as an inner horizon. So, for $0 \leq \nu < 1$ the
canonical ensemble with an equilibrium temperature does not exist,
due to the different Hawking temperatures for the two horizons,
unless they meet at $\nu=1$. Therefore, for $0 \leq \nu <1$ the
black hole decay process in section 3 and the entropy computation
in the section 4 do not apply.

2. Even though I obtained the BH entropy (\ref{SBH}) precisely, by
{\it assuming} (\ref{MM}) for the vacua with the negative mass
eigenvalues $M^{\BTZ}_{ \hbox{\scriptsize min}}=-1/8G$ and $
 M^{\HMTZ}_{\hbox{\scriptsize min}}=-\Theta_{\nu}^2/24G(1+\nu)$, the very
 meaning of (\ref{rr}) for the negative masses, where $r^{\HMTZ}_{+}$ and
 $r_{+}^{\BTZ}$ are pure-imaginary, is not quite clear. But from
 the exact agreement with the BH entropy, the assumed formula
 (\ref{MM}) must have some meaning. A possible explanation of this
 is to consider the black hole decay through the process `$HMTZ\rightarrow
 BTZ \rightarrow BTZ~vacuum$' and the proper value of $M^{\HMTZ}_{\hbox{\scriptsize
 min}}$, which might not be an apparently acceptable state, may be
 extrapolated by the {\it analytic continuation} of (\ref{MM}) due to
 some reasons, such as (\ref{MM}) might have more
 general meaning than (\ref{rr}).

3. The Riemann tensor of the HMTZ metric (\ref{BTZ-scalar}) is
singular at the origin as can be shown by computing the curvature
invariant
\begin{\eq}
\label{RR}
 R^{\mu \nu \al \be}R_{\mu \nu \al \be}=\sqrt{\f{2^{13}}{3}} \f{l
 (GM)^{5/2}}{\sqrt{\nu +1} r^5} +O(r^{-4})
\end{\eq}
near the origin. For $M >0, \nu >-1$ this singularity is hidden
inside the event horizon $r_+$. But, in contrast to the BTZ black
hole, the particle states with $M <0$ have the naked curvature
singularity at the origin; even more, (\ref{RR}), as well as the
metric (\ref{BTZ-scalar}), becomes imaginary as $M$ grows
negative. Moreover, one can easily check that there is $no$
conical singularity at the origin regardless of the sign of $M$.

So, there are some qualitative differences between the AdS$_3$
vacuum of the BTZ black hole and the vacuum (\ref{Mmin}) of the
HMTZ black hole: In contrast to the former, the latter has the
naked curvatures singularity as well as the complex metric and
scalar field. The complex-valuedness might not be a problem if one
$assume$ the decay into the AdS$_3$ vacuum
\footnote{This is not forbidden in principle since both vacua
satisfy the classical Einstein equation, which is required in
identifying the transition amplitude $\Gamma$, due to the absence
of the conical singularities. But, in contrast to $M>0$ case, the
tunnelling $probability$ becomes $|\Gamma|^2 \approx |A|^2$ due to
the imaginary valuedness of $\Delta I_E$.}. However, it is unclear
whether all the massive HMTZ black holes decay into their
corresponding BTZ black holes with the masses given by the
relation (\ref{MM}) before reaching the particle states with
$M<0$, or there are some remnant transient states such as one of
them decays into the AdS$_3$ vacuum.

4. The potential $V_{\nu} (\phi)$ of (\ref{V}) is unbounded below
for $\nu >-1$, such as the weak energy condition (WEC) is violated
for $outside$ the horizon, and the scalar field excitation around
$\phi=0$ becomes ``tachyonic'' with the mass-squared
$m^2=-3/4l^2$. So, the HMTZ black hole departs importantly from
the assumptions of the ``no-scalar-hair'' theorems
\ci{Chas:70,Beke:95}; departs from the old theorems \ci{Chas:70}
due to violation of $\phi dV_{\nu}(\phi)/d \phi \geq 0$ and also
due to the non-vanishing surface integral at infinity, from the
slower fall-off (\ref{phi}); and departs from the new theorem
\ci{Beke:95} due to violation of the $WEC$. \footnote{This seems
to be a generic feature of the hairy solutions. See \ci{Wins:02}.}
However, the decay of the HMTZ into the BTZ realizes $dynamically$
the no-scalar-hair theorems.

5. $V_{\nu} (\phi)$ satisfies the stability bound $m^2 \geq
-1/l^2$ for the perturbations ``on'' AdS$_3$ \ci{Abbo:82}.
\footnote{In the original proof \ci{Abbo:82}, $d \geq 4$ is
implicitly assumed. But this extends also to $d=3$.} But this does
not necessarily guarantees the stability of the HMTZ solution
since this solution is not ``globally'' AdS$_3$: It is already
known, due to Mart\'{i}nez, that $\nu=0$ solution is $unstable$
under linear perturbations of the metric \ci{Mart:98}; I suspect
that $\nu >0$ solutions would be also unstable since this is
``more'' tachyonic than the $\nu=0$ solution, which is unstable
already; $-1< \nu <0$ solutions ``might'' be stable though it is
still tachyonic; an explicit computation on this stability under
linear perturbations would be certainly interesting.

6. There is the mirror black hole solution with $\tilde{M}=-M,
1+\tilde{\nu}=-(1+\nu) <0$ ($\tilde{\nu}=\nu=-1$ solution
corresponds to the massless black holes) with the metric
$\tilde{g}_{\mu \nu}=g_{\mu \nu}$ and the scalar field
$\tilde{\phi}=\phi$ [$\tilde{B}=B, \tilde{H}=H,
\Theta_{\tilde{\nu}}=\Theta, \tilde{r}_{+}=r_+, \tilde{T}=T$] for
\begin{\eq}
&&\tilde{V}_{\tilde{\nu} }(\tilde{\phi} )=-\frac{1}{8l^{2}}\left(
\cosh ^{6}\tilde{\phi} -(\tilde{\nu}+2) \sinh
^{6}\tilde{\phi} \right)=V_{\nu}(\phi), \\
&&\tilde{F}=\frac{\tilde{H}^{2}}{l^{2}}+(1+\tilde{\nu} )\left(
\frac{3\tilde{B}^{2}}{l^{2}}+\frac{2\tilde{B}^{3}}{
l^{2}\tilde{H}}\right)=F;
\end{\eq}
one can not distinguish this with the HMTZ black hole except for
$\tilde{M}<0, \tilde{\nu}<-1$. The role of this mirror solution to
the vacuum decay into the AdS$_3$ vacuum
is not clear.

7. The higher order corrections in the bulk, due to the thermal
fluctuations of black hole geometry--fluctuations of
the metric-- to the BH entropy is given by, following
\ci{Park:04},
\begin{\eq}
\label{Sbulk} S=S_{\BH}-\f{3}{2} \mbox{ln} S_{\BH} +\f{1}{2}
\mbox{ln}\left[\f{ \pi^3 l^4 (\de E)^2   }{2 G^2}
\left(\frac{3(1+\nu )}{ \Theta _{\nu }^2 } \right)^2\right] ~.
\end{\eq}
On the other hand, the corresponding correction to statistical
entropy $a'~la$ Strominger is given by
\begin{\eq}
S=S_{\BH}-3 ~\mbox{ln} S_{\BH} +\mbox{ln}\left[\f{64 \pi^3 l^2
}{(8 G)^{2}}  \frac{ \Theta _{\nu } }{\sqrt{3(1+\nu )}}  \right]~.
\end{\eq}
This result shows the factor of 2 disagreement with the bulk
result (\ref{Sbulk}) as in the BTZ black hole without a scalar
field \ci{Park:04,Myun:04}. It would be interesting to compute the
contribution due to the fluctuation of the scalar field. It would
be also interesting to compare this with that of the horizon
holography $a'~la$ Carlip \ci{Carl:99b,Park:99b,Park:04,Carl:02}.

8. My result implies that the CFT for the HMTZ black hole has no
$SL(2, \bf{C})$ invariant vacuum, which generates zero-eigenvalues
for the generators $\Delta$ and $\bar{\Delta}$. This is hard to
understand from standard CFTs \ci{Card:86}. But such an issue is
always the case for the Strominger's method \ci{Carl:98} and needs
further investigation\footnote{For some recent discussions, see
also Ref. \ci{Fjel:02}.}.

\begin{center} {\bf Acknowledgments}
\end{center}

I appreciate Dr. Gungwon Kang for remind me of Ref. \ci{Nats:99}
and related discussions. I also appreciate Dr. Cristian
Mart\'{i}nez for kindly explaining his computation and
illuminating discussions. This work was supported by the Korean
Research Foundation Grant (KRF-2002-070-C00022).

\end{document}